\documentclass[12pt,a4paper]{article}
\usepackage{amssymb}

\usepackage{cite}
\newlength{\HFPP}       \HFPP5.4mm
\makeatletter
\@addtoreset{equation}{section}
\makeatother

\newcommand{\one}[1]{\mathop{#1}\limits^1}
\newcommand{\two}[1]{\mathop{#1}\limits^2}

\begin{document}

%%\begin{titlepage}
\def\thefootnote{\fnsymbol{footnote}}
\def\preprint#1#2{\noindent\hbox{#1}\hfill\hbox{#2}\vskip 10pt}

\preprint{ITP-UH-05/99}{March 1999}
\vfill

\begin{center}
  {\Large\sc Open $t$-$J$ chain with boundary impurities}

%% Author(s):
%%
\vfill

{\sc Gerald Bed\"urftig}\footnote{e-mail: bed@itp.uni-hannover.de} and 
{\sc Holger Frahm}\footnote{e-mail: frahm@itp.uni-hannover.de} 
\vspace{1.0em}

%% Address:
%%
{\sl
  Institut f\"ur Theoretische Physik, Universit\"at Hannover,
  D-30167~Hannover, Germany}\\
\end{center}
\vfill

\begin{quote}
We study integrable boundary conditions for the supersymmetric
$t$--$J$ model of correlated electrons which arise when combining
static scattering potentials with dynamical impurities carrying an
internal degree of freedom.  The latter differ from the bulk sites by
allowing for double occupation of the local orbitals.  The spectrum of
the resulting Hamiltonians is obtained by means of the algebraic Bethe
Ansatz.
\end{quote}

\vfill

%{PACS-Nos.: 
%71.10.Fd,	% Lattice fermion models (Hubbard model, etc.)
%71.10.Pm,	% Fermions in reduced dimensions (anyons, composite fermions,
%         	% Luttinger liquid, etc.)
%73.20.-r	% Surface and interface electron states
%}
%
\vspace*{\fill}
\setcounter{footnote}{0}
%%\end{titlepage}

%\begin{multicols}{2}
%\narrowtext 
%
%
\section{Introduction}

Impurities in correlated quantum systems have attracted considerable
interest recently.  In particular in one spatial dimension exactly
solvable models and powerful field theoretical methods have provided
insights into the properties of local perturbations of ideal chain
systems
\cite{hschulz:87,gohs:90,kafi:92,soea:93,assu:96b,affl:97,esfr:97}.
Static perturbations such as scattering potentials have a profound
effect on the transport properties of quasi one-dimensional structures
such as quantum wires.  Impurities with internal degrees of freedom,
e.g. a localized magnetic moment in the Kondo problem, may be screened
due to resonances with the electrons in the 1d correlated host.

In the framework of the Quantum Inverse Scattering Method (QISM)
\cite{vladb} the construction of integrable models for such systems is
based on inhomogeneous vertex models constructed from solutions to a
Yang-Baxter equation.  Such inhomogeneities have first been embedded
into periodic chains by Andrei and Johannesson for the Heisenberg
model \cite{anjo:84} and later into various models including the
supersymmetric $t$--$J$ model of interacting electrons by various
authors \cite{bares:95,bef:96,bef:97,sczv:97}.  A direct consequence
of this way of construction is the lack of backscattering at the
impurities \cite{ecpr:97}.  Consideration of such impurities in a more
general field theoretical approach has led to the conclusion that the
interactions in the integrable models are fine tuned to a fixed point
which is unstable under renormalization flow \cite{soea:93}.  As a
consequence the integrable inhomogeneities lack characteristic
properties of a generic potential scatterer in a 1d system with
repulsive interactions, which has been found to drive the open chain
fixed point leading to a vanishing of the conductivity \cite{kafi:92}.

This can be overcome by combination of these integrable
inhomogeneities with a real boundary.  Again the construction of such
models is possible within the QISM from solutions to the reflection
equations (RE) \cite{cher:84,skly:88} imposing consistency conditions
on the possible boundary conditions for a given bulk system.  For the
$t$--$J$ model the simplest such ($c$-number) solutions of the RE
correspond to boundary chemical potentials and boundary magnetic
fields respectively \cite{gonz:94,essl:96}.  Combining these boundary
matrices with solutions of the Yang-Baxter equation one can derive
dynamical boundary impurity models: Heisenberg models
\cite{frzv:97b,wang:97} and Kondo type impurities
\cite{hschulz:87,Wang97,Huxx98} coupled to correlated electron systems
have been studied this way.

Here we construct the most general boundary impurities that
can be realized within this approach by combination of the known
static boundary fields for the supersymmetric $t$--$J$ model with a
dynamical impurity allowing for double occupancy of the electronic
orbital at its site.  This four state impurity alone has been studied
previously for periodic chains and open ones with reflecting ends
\cite{bef:96,bef:97,bfz:up}.  The resulting boundary terms are
characterized by the boundary field and in addition by a real parameter
characterizing the four dimensional typical representation of the
graded Lie algebra $gl(2|1)$ realized on the Hilbert space of the
impurity and by its coupling strength to the host system which is
controlled by a shift in the spectral parameter of the corresponding
vertex. Using both boundary chemical potentials and boundary magnetic
fields this leads to two three parametric families of boundary terms.
Further models are obtained by application of the 'projecting method'
introduced recently \cite{frsl:99}.  Finally, we present the Bethe
Ansatz equations determining the spectra of these impurity models.

%%%%%%%%%%%%%%%%%%%%%%%%%%%%%%%%%%%%%%%%%%%%%%%%%%%%%%%%%%%%%%%%%%%%%%
\section{Algebraic Construction}
Following \cite{cher:84,skly:88} the classification of integrable
boundary conditions within the QISM is based on representations of two
algebras ${\cal T}_\pm$ defined in terms of reflection equations (RE).
For ${\cal T}_-$ this equation reads:
\begin{equation}
  {\cal R}^{12}(\lambda-\mu)
      \one{{\cal T}_-}(\lambda){\cal R}^{21}(\lambda+\mu)
	\two{{\cal T}_-}(\mu)
 =\two{{\cal T}_-}(\mu){\cal R}^{12}(\lambda+\mu)
      \one{{\cal T}_-}(\lambda){\cal R}^{21}(\lambda-\mu)\ ,
\label{eq:ref}
\end{equation}
with $\one{{\cal T}_-}= {\cal T} \otimes I$ and $\two{{\cal T}_-}= I
\otimes {\cal T}$.\footnote{For the $t$--$J$ model considered here
these tensor products carry a grading and we have to use a graded
version of the QISM. For details see for example \cite{esko:92}.}  The
algebra ${\cal T}_+$ is related to ${\cal T}_-$ by an automorphism.
Representations of ${\cal T}_\pm$ determine the boundary terms in the
Hamiltonian at the left (right) end of the chain.  Since these can be
chosen independently it is sufficient to consider solutions of (2.1)
to obtain a classification of the possible boundary impurities.

The matrix ${\cal R}$ in (\ref{eq:ref}) solves the 
quantum Yang-Baxter equation (YBE):
\begin{equation}
{\cal R}^{12}(\lambda){\cal R}^{13}(\lambda+\mu){\cal R}^{23}(\mu)=
{\cal R}^{23}(\mu){\cal R}^{13}(\lambda+\mu){\cal R}^{12}(\lambda)
\end{equation}
where the superscripts denote the spaces in the tensor product
$V_1\otimes V_2\otimes V_3$ in which ${\cal R}^{ij}$ acts
nontrivially.

For the  $t$--$J$ model this ${\cal R}$-matrix is given by:
\begin{equation}
  \left({\cal R}^{12}(\lambda)\right)_{i_1i_2}^{j_1j_2}
  =\frac{\lambda}{\lambda+i}\, \delta_{i_1}^{j_1}\delta_{i_2}^{j_2}  +
   \frac{i}{\lambda+i}\, \Pi_{i_1i_2}^{j_1j_2} \ , 
\label{eq:Rtj}
\end{equation}
with the graded permutation operator $\Pi_{ab}^{cd}=\delta_a^d
\delta_c^b (-1)^{[a][b]}$, $[a]\in\{0,1\}$ denoting the grading of the
basis states.  The $c$-number solutions of the RE (\ref{eq:ref})
corresponding to this ${\cal R}$-matrix can be classified
\cite{gonz:94}, below we shall use the diagonal ones
\begin{equation}
  K_-^p=\left(\begin{array}{ccc}
	1 && \\
	& 1 & \\
	&& -{{p\lambda+{i}} \over {p\lambda-{i}}} 
	\end{array}\right), \quad
  K_-^h=\left( \begin{array}{ccc}
	-{{h\lambda+{i}} \over {h\lambda-{i}}} && \\
	& 1 & \\
	&& 1 
	\end{array}\right)
\label{kdiag}
\end{equation}
corresponding to a boundary chemical potential $p$ and a boundary
magnetic field $h$ (in combination with a chemical potential),
respectively \cite{essl:96}.

To construct boundary impurities carrying internal degrees of freedom
we combine the matrices from (\ref{kdiag}) with an integrable impurity
which has been considered previously in a periodic chain
\cite{bef:96,bef:97}.  In the QISM this impurity is characterized by the
following ${\cal L}$-matrix:
\begin{equation}
   {\cal L}^{34}(\lambda)=
   \frac{\lambda-i({\alpha \over 2}+1)}{\lambda+i({\alpha \over 2}+1)}
   +\frac{i}{\lambda+i({\alpha \over 2}+1)}\tilde{\cal L}\ ,
\quad \tilde{\cal L}= 
  \left( \begin{array}{ccc} 
  1-n_{\uparrow} &  - S^- &   Q_{\uparrow} \\
  - S^+ &   1-n_{\downarrow} &    Q_{\downarrow} \\
    Q_{\uparrow}^\dagger&   Q_{\downarrow}^\dagger& \alpha+ 2-n\\
  \end{array} \right) \ .
\label{lop4}
\end{equation}
Here $n=\sum_{\sigma=\uparrow,\downarrow} n_{\sigma} = \sum_\sigma
c^{\dagger}_{\sigma} c_{\sigma}$ and $\vec{S}=\frac{1}{2}
c^\dagger_{\alpha} \vec{\sigma}_{\alpha\beta} c_{\beta}$ are the
electron number and spin operators on the impurity site expressed in
terms of canonical fermionic creation and annihilation operators.  The
$Q_{\sigma}$ are the fermionic generators of $gl(2|1)$ in this
representation which can be expressed in terms of projection operators
(the so called `Hubbard Operators') ${\bf X}^{ab}=|a\rangle\langle b|$
with $a,b=\uparrow,\downarrow,2,0$:
\begin{equation}   
  Q_{\sigma}=\sqrt{\alpha+1}{\bf X}^{0 \sigma}
   -2\sigma \sqrt{\alpha} {\bf X}^{-\sigma 2} \ 
\end{equation}
with $\sigma=\pm {1 \over 2}$ corresponding to $\sigma=\uparrow,\downarrow$.\\ 
${\cal L}^{34}$ acts on a four-dimensional quantum space and satisfies 
the intertwining relation:
\begin{equation}
  {\cal R}^{12}(\lambda-\mu)
	\left({\cal L}^{34}(\lambda)\otimes{\cal L}^{34}(\mu)\right)=
  \left({\cal L}^{34}(\mu)\otimes{\cal L}^{34}(\lambda)\right)
	{\cal R}^{12}(\lambda-\mu)\ .
\label{eq:ir}
\end{equation}
Following Sklyanin \cite{skly:88} operator-valued matrix solutions of
the RE are obtained by ``dressing'' a c-number solution $K_-(\lambda)$
of the RE with ${\cal L}^{34}$, i.e.\ considering the product ${\cal
T}_-(\lambda)= {\cal L}^{34}(\lambda+t)K_-(\lambda) \left({\cal
L}^{34}(-\lambda+t) \right)^{-1}$ in matrix space (a shift $t$ of the
spectral parameter $\lambda$ is consistent with the intertwining
relation (\ref{eq:ir})).  To construct an integrable chain with this
impurity placed on site $1$ this reasoning is iterated with the ${\cal
L}$-operators for the $t$--$J$ model, i.e.\ ${\cal L}_n={\cal
R}^{0n}$, $n=2,\ldots,L$ resulting in
\begin{equation}
  {\cal T}_-(\lambda)= {\cal L}_L(\lambda) \cdots {\cal L}_2(\lambda)
  {\cal L}^{34}_1(\lambda+t) K_-(\lambda)
  \left({\cal L}^{34}_1(-\lambda+t)\right)^{-1}
  \left({\cal L}_2(-\lambda)\right)^{-1}\cdots
  \left({\cal L}_L(\lambda)\right)^{-1} \ .
\end{equation}
The integrable model is now defined through the transfer matrix
\begin{equation}
  \tau(\lambda)=\mbox{str}_0
  \left[K_+(\lambda){\cal T}_-(\lambda)\right]\ .
\end{equation}
$\mbox{str}_0(M)=\sum_a (-1)^{[a]}M_{aa}$ is the (graded) supertrace
taken in matrix space.  Since the purpose of this paper is the
classification of integrable boundary terms obtained in this class we
restrict ourselves to the simplest case of $K_+(\lambda)\equiv
\mathbf{1}$ as a representation of the algebra ${\cal T}_+$ which
corresponds to a reflecting left boundary of the chain.  Then the
Hamiltonian is obtained by differentiation of the transfer matrix with
respect to the spectral parameter:
\begin{equation} 
  H \propto i{\partial \over \partial\lambda} 
	\tau(\lambda)|_{\lambda=0}\ . 
\end{equation} 
This leads to the following Hamiltonian of the quantum chain
\begin{eqnarray}
  {\cal H}&=& -{\cal P}\left(\sum_{j=2}^{L-1}
	\sum_\sigma c^\dagger_{j,\sigma}c_{j+1,\sigma}
	+c^\dagger_{j+1,\sigma}c_{j,\sigma}\right){\cal P}
\nonumber\\
  &&+2\sum_{j=2}^{L-1} \left[ {\vec S_j}{\vec S_{j+1}}
	-\frac{n_jn_{j+1}}{4} +{1 \over 2} (n_j+n_{j+1}) \right]  
	+{4 \over 4 t^2+({\alpha}+2)^2} {\cal H}_{b}^{p,h}\ ,
\label{hamil}
\end{eqnarray}
where ${\cal P}$ projects out double occupancies on the bulk sites,
and $\vec{S_j}$, $n_j$ are the electronic spin and number operators on
site $j$ defined as above. The boundary terms ${\cal H}_b$ depend on
the choice of the boundary matrix, after a unitary transformation
${\cal H}_{b}^p$ is given by
\begin{eqnarray}
{\cal H}_b^p=&-&p\left( t^2+{(\alpha+2)^2 \over 4} \right)
  +\left\{1+p(\alpha+1)\vphantom{a^2 \over b}\right\}{n_1}
  +\left\{1+\alpha+p\left(t^2+{\alpha^2 \over 4}\right)\right\}{n_2}
\vphantom{ \sqrt{(\alpha+1) \left(\left[1+{p \alpha \over 2}\right]^2 \right)}} \nonumber\\
&+& \left\{1+p \alpha\vphantom{\vec S_1}\right\}
	\left\{2 {\vec S_1}{\vec S_{2}}-\frac{n_1n_{2}}{2}\right\}
 +p\left\{n_2-2 \vphantom{\vec S_1} \right\} {\bf X}_1^{22} \nonumber\\
&-& \sqrt{\alpha+1}\,t_0\left\{{\bf X}_2^{\uparrow 0}{\bf X}_1^{0 \uparrow} 
	+ {\bf X}_2^{\downarrow 0}{\bf X}_1^{0
	\downarrow}+h.c.\right\}
\label{hamilbo} \\
&-& \sqrt{\alpha}\,t_2  
  \left\{{\bf X}_2^{\uparrow 0}{\bf X}_1^{\downarrow 2} - 
  {\bf X}_2^{\downarrow 0}{\bf X}_1^{\uparrow 2}+h.c.\right\}  
\nonumber \ ,
\end{eqnarray}
where $t_0=\sqrt{\left(\left[1+{p \alpha \over
2}\right]^2+p^2t^2\right)}$ and $t_2=\sqrt{\left( \left[1+p
\left({\alpha \over 2}-1\right)\right]^2+p^2t^2\right)}$.  Note that
the representation of $gl(2|1)$ entering (\ref{lop4}) does allow for
double occupancy on the first site.

Similarly, we obtain the following boundary operator when considering
the ${K}^h_-$-matrix:
\begin{eqnarray}
{\cal H}_b^h&=& {n_1}+(\alpha+1){n_2}-h(1+\alpha){n_{1\uparrow}}
   -h \left(t^2+{\alpha^2 \over 4}\right)n_{2\uparrow} 
\nonumber\\
&& +(\alpha h -1) n_{1\uparrow} n_{2\downarrow}
  - n_{1\downarrow} n_{2\uparrow}
+h\left(1-n_{2\uparrow}\right) {\bf X}_1^{22} \nonumber\\
&& +\sqrt{\alpha}(1+h)\left\{{\bf X}_2^{\downarrow 0}{\bf X}_1^{\uparrow 2}
	+h.c.\right\}
    -\sqrt{\alpha+1}\left\{{\bf X}_2^{\downarrow 0}{\bf X}_1^{0 \downarrow}
	+h.c.\right\} \\
\label{HamH}
&& -t_c\left[
    \sqrt{\alpha}\,{\bf X}_2^{\uparrow 0}{\bf X}_1^{\downarrow 2}
    +\sqrt{\alpha+1}\,{\bf X}_2^{\uparrow 0}{\bf X}_1^{0\uparrow}
  +S_1^+S_2^-+h.c.\right] \nonumber 
\end{eqnarray}
with $t_c=\sqrt{\left(1-{\alpha h \over 2}\right)^2+t^2h^2}$.

The models constructed above can be solved using the algebraic Bethe
Ansatz.  Starting from the completely filled, fully polarized state,
i.e.\ doubly occupied impurity site $1$ and all other sites are
occupied by a spin-$\uparrow$ electron, we find that the spectrum of
${\cal H}$ is determined by the solutions of the Bethe Ansatz
equations (BAE)
\begin{eqnarray}
 &&B_h e_1^{2(L-1)}(\lambda_k) =\prod_{j\neq k}^{M_s}
	e_2(\lambda_k-\lambda_j) e_2(\lambda_k+\lambda_j) 
	\prod_{\ell=1}^{M_c} e_{-1}(\lambda_k-\vartheta_\ell)
	e_{-1}(\lambda_k+\vartheta_\ell)\nonumber\\
 &&\prod_{j=1}^{M_s}e_{-1}(\vartheta_\ell-\lambda_j)
	e_{-1}(\vartheta_\ell+\lambda_j)=
	B_p(\vartheta_\ell)e_{\alpha}(\vartheta_\ell+t)
	e_{\alpha}(\vartheta_\ell-t)\ ,
\label{bae}
\end{eqnarray}
where $e_n(x)=\frac{x+{in}/{2}}{x-{in}/{2}}$, $M_c=L+1-N_e$, 
$M_s=L-N_\uparrow$ and boundary phase shifts
\begin{equation}
  B_h(\lambda)=\left\{ 
	\begin{array}{cl}
	1 & \mbox{for } {K}^{^p}_- \\
	-e_{-1-{2 \over h}}(\lambda)  & \mbox{for } {K}^{h}_-
	\end{array}\right. \quad \mbox{and} \quad
  B_p(\vartheta)=\left\{
	\begin{array}{cc}
	-e_{{2 \over p}-2}(\vartheta) & \mbox{for } {K}^{^p}_- \\
	1 & \mbox{for } {K}^{h}_-
	\end{array}
	\right. .
\end{equation}
The energy of the corresponding Bethe state
% with spectral parameters $\{\lambda_j\}$ and $\{\vartheta_\ell\}$ 
is then given by the expression
\begin{equation}
  E=E_b^{p,h}+2(L-2)-
  \sum_{j=1}^{M_s}\frac{1}{\lambda_j^2+\frac{1}{4}}
%%+\left(\mu-{H \over 2}\right)M_c+HM_s-\mu(L+1)-{H \over 2}(L-1)
\end{equation}
with $E_b^{p}=(4 \alpha+8)/(4t^2+(\alpha+2)^2)$ and $E^h_b=\left(4
\alpha+8\right)/\left(4t^2+(\alpha+2)^2\right)-h$.

As for the periodic and the open $t$--$J$ model the zero temperature
ground state and the low--lying charged and magnetic excitations are
characterized by real solutions for the $\lambda$ and $\vartheta$
rapidities of the BAE.

%%%%%%%%%%%%%%%%%%%%%%%%%%%%%%%%%%%%%%%%%%%%%%%%%%%%%%%%%%%%%%%%%%%%%%
\section{Projecting method}
Recently, it has been realized that new integrable boundary
Hamiltonians may be obtained after fine tuning of the parameters
characterizing the boundary and impurity, respectively \cite{frsl:99}
by projection onto an invariant subspace.  An important application of
this procedure is a Kondo spin coupled to the $t$--$J$ model
\cite{Wang97,Huxx98,Zhou98.2}.
To apply the projecting method one has to find a decomposition of the
Hilbert space of the impurity ${\sf H}={\sf H}_1 \oplus {\sf H}_2$
and fine tune the parameters in ${\cal T}_-$ such that one of the
following conditions is satisfied
($\Pi_{1,2}$ are projectors onto ${\sf H}_{1,2}$)
\begin{equation}
  \Pi_1{\cal T}_- \Pi_2 =0 \quad 
  \mbox{or} \quad \Pi_2{\cal T}_- \Pi_1 =0 \ .
\label{eq:proj}
\end{equation}
Starting from ${K}^{p}_-$ we find that the decomposition ${\sf
H}=\{\uparrow,\downarrow,0\}\oplus \{2\}$ is possible for $t=i\left(
{\alpha \over 2}-1+{1 \over p}\right)$, the resulting Hamiltonian do
not lead to new models.  For ${\cal H}_2(\alpha,p)$ the impurity is a
scalar one and the model reduces to an open $t$-$J$ model with
boundary chemical potential at site two.  The other possible model,
${\cal H}_{\uparrow,\downarrow,0}(\alpha,p)$, turns out to be a
simple reparametrisation of ${\cal H}_{\uparrow,\downarrow,0}
(\alpha=0,p,t)$.

Choosing $t=i\left( {\alpha \over 2}+{1 \over p}\right)$ the condition
(\ref{eq:proj}) is satisfied for the decomposition ${\sf H}=\{2,
\uparrow,\downarrow\}\oplus \{0\}$. Again ${\cal H}_0(\alpha,p)$ is
just an open $t$-$J$ model with boundary chemical potential at the
second site.  On the second subspace, however, we find a
two-parametric Hamiltonian:
\begin{eqnarray}
{\cal H}_{2, \uparrow,\downarrow}=&-&(a+b+2) 
  +{(a+b+2)^2 \over a+b+1}\,{n_1}
  +{a+b+2 \over 3+2a+b}\, {n_2}
\nonumber\\
&+& {(a+1)(a+b+2)^2 \over (a+b+1)(3+2a+b)}\,
  \left\{{ 2 {\vec S_1}{\vec S_{2}}}-\frac{n_1n_{2}}{2}\right\}
\nonumber\\
&+&{(a+b+2)^3 \over(a+b+1)(3+2a+b)}\,
  \left\{n_2-2 \vphantom{\vec S_1} \right\} {\bf X}_1^{22}
\nonumber\\
&-& {(a+b+2)^2 \over(a+b+1)(3+2a+b)}\,
\sqrt{ab}
\left\{{\bf X}_2^{\uparrow 0}{\bf X}_1^{\downarrow 2} 
 - {\bf X}_2^{\downarrow 0}{\bf X}_1^{\uparrow 2}+h.c.\right\}  
\nonumber \ .
\label{hamilbop2}
\end{eqnarray}
The Hamiltonian ${\cal H}_{2, \uparrow,\downarrow}$ may be constructed 
with aid of ${\cal L}^{34}(\alpha=-1)$ and choosing the remaining two
parameters as $\tilde{t}=-{i \over 2} {(a+b)/(a+b+2)}$ and 
$\tilde{p}={(a+b+2)/(3+2a+b)}$. This ${\cal L}$-matrix corresponds 
to the one for the dual space of the fundamental three dimensional
representation used in \cite{abri:98,lifo:99} to construct periodic 
$t$-$J$ models with alternating impurities and usual $t$-$J$ sites.
Choosing $a=0$ or $b=0$ a further projection is possible:
${\sf H}=\{2\} \oplus \{\uparrow,\downarrow\}$. ${\cal H}_{2}$ corresponds 
to a boundary chemical potential at the second site. Substituting 
$b=-{ a \over 1+a}$ in ${\cal H}_{\uparrow,\downarrow}^{a=0}$ one obtains 
${\cal H}_{\uparrow,\downarrow}^{b=0}$. The resulting Hamiltonian 
${\cal H}_{\uparrow,\downarrow}^{a=0}$ can be identified with a 
Spin-${1 \over 2}$ Kondo impurity introduced in \cite{Wang97}. 
 
Considering ${K}^{h}_-$ only one decomposition ${\sf H}=\{2,\uparrow\}
\oplus \{0,\downarrow\}$ satisfying (\ref{eq:proj}) is possible.
Choosing $t=i\left( {\alpha \over 2}-{1 \over h}\right)$ we find two
new boundary Hamiltonians, namely
\begin{eqnarray}
  {\cal H}_{2,\uparrow}&=&\left[ {h^2 \over (h+1)(h(\alpha+1)-1)}\right]
  \left\{ {n_1}+(\alpha+1)\left({n_2}-hn_{1\uparrow}\right)
  +n_{2\uparrow} {1- \alpha h \over h}
 \right. 
\nonumber\\
&+&\left. \vphantom{{1- \alpha h \over h}}
  (\alpha h -1)n_{1\uparrow} n_{2\downarrow}+
  h\left(1-n_{2\uparrow}\right) {\bf X}_1^{22}\right\} 
  +{\sqrt{\alpha}h^2 \over h(\alpha+1)-1}\left\{{\bf X}_2^{\downarrow 0}
   {\bf X}_1^{\uparrow 2}+h.c.\right\} 
\label{Ham2u}
\end{eqnarray}
and 
\begin{eqnarray}
{\cal H}_{\downarrow,0}&=&\left[ {h^2 \over (h+1)(h(\alpha+1)-1)}\right]
\left\{ {n_1}+(\alpha+1){n_2}
  +{1- \alpha h \over h}n_{2\uparrow} 
  -n_{1\downarrow} n_{2\uparrow}  \right.
\nonumber\\
 && \qquad\left.\vphantom{{1- \alpha h \over h}}
  -\sqrt{\alpha+1}\left\{{\bf X}_2^{\downarrow 0}{\bf X}_1^{0
  \downarrow}
+h.c.\right\} 
\right\}\ .
\label{Hamd0}
\end{eqnarray}
The BAE for the projected Hamiltonians coincide with the ones for the
original model (\ref{bae}), \emph{provided} that the reference state
used in their construction, i.e.\ the state $|2\rangle$ in the impurity
Hilbert space is not projected out.  Hence, to obtain the spectrum of
${\cal H}_{\downarrow,0}$ one has to use different BAE obtained for a
suitable pseudo-vacuum.  Alternatively, one may consider solutions of
(\ref{bae}) \emph{after} adding the complex solutions corresponding to
bound states  (see e.g.\ Refs.~\cite{esfr:97,befr:97}):
\begin{equation}
  \lambda=i \left({1\over 2}+{1 \over h}\right) 
   \quad \mbox{and} \quad \vartheta = {i \over h} \ .
\end{equation}
This results in the following set of BAE for for the Hamiltonian
(\ref{Hamd0}) 
\begin{eqnarray}
 -e_{3+2/h}(\lambda_k)e^{2(L-1)}_1(\lambda_k)&=&\prod_{j\neq k}^{M_s}
 e_2(\lambda_k-\lambda_j) e_2(\lambda_k+\lambda_j) 
 \prod_{\ell=1}^{M_c} e_{-1}(\lambda_k-\vartheta_\ell)
 e_{-1}(\lambda_k+\vartheta_\ell)\nonumber\\
 \prod_{j=1}^{M_s}e_{-1}(\vartheta_\ell-\lambda_j)
 e_{-1}(\vartheta_\ell+\lambda_j)&=&  
 e_{2\alpha-2/h}(\vartheta_\ell)e_{2+2/h}((\vartheta_\ell)\ ,
\label{baeh}
\end{eqnarray}
where $M_c=L-N_e$ and $M_s=L-1-N_\uparrow$.  The energy of the
corresponding Bethe state with spectral parameters $\{\lambda_j\}$ and
$\{\vartheta_\ell\}$ is given by
\begin{equation}
E={h \over h (\alpha +1) -1}+2(L-2)-
\sum_{j=1}^{M_s}\frac{1}{\lambda_j^2+\frac{1}{4}}
%%+\left(\mu-{H \over 2}\right)M_c+HM_s-\mu(L+1)-{H \over 2}(L-1)
\end{equation}

%%%%%%%%%%%%%%%%%%%%%%%%%%%%%%%%%%%%%%%%%%%%%%%%%%%%%%%%%%%%%%%%%%%%%%
\section{Summary}
Starting from a particular solution of the intertwining relation
(\ref{eq:ir}) built from a four-dimensional representation of the
graded Lie algebra $gl(2|1)$ and diagonal $c$-number solutions
(\ref{kdiag}) of the RE we have constructed supersymmetric $t$--$J$
models with integrable boundary impurities of Anderson or Kondo type,
i.e.\ with an internal degree of freedom.  Due to non-zero boundary
potentials the resulting boundary terms break the supersymmetry of the
model, the most general ones which can be constructed this way are
given in Eqs.~(\ref{hamilbo}) and (\ref{HamH}).  The presence of the
boundary allowed for fine tuning of these potentials to project the
model to a remaining invariant subspace.  In most cases this
projection led to models which had been constructed directly before.
Eqs.\ ({\ref{Ham2u}) and (\ref{Hamd0}) however, are novel.

The spectra of these models have been obtained by means of the
algebraic Bethe Ansatz.  Furthermore, having the exact dependence of
the ground state energy on the parameters defining the impurity allows
for the computation of \emph{local} correlations which usually are not
easily accessible from the Bethe Ansatz solution (see Refs.\
\cite{bef:97,frzv:97b} for examples).  

%%%%%%%%%%%%%%%%%%%%%%%%%%%%%%%%%%%%%%%%%%%%%%%%%%%%%%%%%%%%%%%%%%%%%%
\section*{Acknowledgements}
We would like to thank A.\,A.\ Zvyagin for discussions.  This work has
been supported by the Deutsche Forschungsgemeinschaft under Grant No.\
Fr~737/2--3.

\setlength{\baselineskip}{13pt}

\end{document}